\documentclass{jfm}

\usepackage{slashbox}
\usepackage[dvips]{graphicx}
\usepackage{natbib}
\usepackage{bm}
\usepackage[normalem]{ulem}
\usepackage[usenames]{color}
\usepackage{psfrag}
\usepackage{textcomp}
\usepackage{amsmath}

\newcommand\Rey{\mbox{\textrm{Re}}}  
\newcommand\Pra{\mbox{\textrm{Pr}}} 
\newcommand\Ra{\mbox{\textrm{Ra}}} 
\newcommand\Nu{\mbox{\textrm{Nu}}}

\begin{document}

\title[Turbulent RBC in a tilted $\Gamma=0.50$ cylinder]{Effect of tilting on turbulent convection: Cylindrical samples with aspect ratio $\Gamma=0.50$ }

\author{Stephan Weiss\thanks{Current address: Department of Physics \& Center for the Study of Complex Systems, University of Michigan, Ann Arbor, MI 48109 USA
 } and Guenter Ahlers}
\affiliation{Department of Physics, University of California, Santa Barbara, CA 93106, USA}

\maketitle

\begin{abstract}
We report measurements of properties of turbulent thermal convection of a fluid with a Prandtl number $\Pra=4.38$ in a cylindrical cell with an aspect ratio $\Gamma=0.50$. The rotational symmetry was broken by a small tilt of the sample axis relative to gravity. 
Measurements of the heat transport (as expressed by the Nusselt number \Nu), as well as of large-scale-circulation (LSC) properties by means of temperature measurements along the sidewall, are presented.
In contradistinction to similar experiments using containers of aspect ratio $\Gamma=1.00$ \cite[]{ABN06} and $\Gamma=0.50$ \cite[]{CRCC04,SXX05,RGKS10}, we see a very small  increase of the heat transport  for tilt angles up to about 0.1 rad. 
Based on measurements of properties of the LSC we explain this increase by a stabilization of the single-roll state (SRS) of the LSC and a de-stabilization of the double-roll state (DRS) (it is known from previous work that  the SRS has a slightly larger heat transport than the DRS).
Further, we present quantitative measurements of the strength of the LSC, its orientation, and its torsional oscillation as a function of the tilt angle.
\end{abstract}

\section{Introduction}

Turbulent fluid motion driven by a temperature gradient plays an important role in a variety of natural and industrial processes. Thus, it has been a topic of interest for many decades [for  reviews intended for a broad audience, see \cite{Ka01,Ah09}; for more specialized reviews, see \cite{AGL09,LX10}]. It is relevant to fundamental problems in astro- or geophysics \cite[]{HL80b,Ra00}, but plays an important role also in applications ranging from the optimization of heat distribution inside aircraft cabins \cite[]{KBW09} to the cooling of giant nuclear reactors \cite[]{Sta10}. 

In order to understand the fundamental processes of turbulent thermal convection, experiments and simulations focus on simple systems with well defined boundaries known as Rayleigh-B\'enard convection (RBC). Such systems usually consist of a Newtonian fluid that is confined by a warm horizontal plate from the bottom and a colder parallel plate at a distance $L$ from the top.
When the temperature difference $\Delta T$ between the bottom ($T_b$) and the top plate ($T_t$) is not too large, {\em i.e.} when the fluid properties do not change significantly within that temperature range, the Oberbeck-Boussinesq approximation \cite[]{Ob79,Bo03} can be applied and for a given geometry the state of the system is defined by only two dimensionless parameters.
These are the ratio of the driving buoyancy to the viscous and thermal damping forces which is expressed by the Rayleigh number
\begin{equation}
\Ra=\frac{g\alpha\Delta T L^3}{\kappa \nu}\mbox{,}
\end{equation}
and the ratio of the kinematic viscosity $\nu$ to the thermal diffusivity $\kappa$ which is given by the Prandtl number
\begin{equation}
\Pra=\frac{\nu}{\kappa}\mbox{.}
\end{equation}
Here, $g$ and $\alpha$ denote the gravitational acceleration and the isobaric thermal expansion coefficient.

A major issue of interest in RBC is the heat transport between the bottom and the top that is expressed  in terms of a dimensionless parameter known as the Nusselt number
\begin{equation}\label{eq:Nu}
\Nu=\frac{\lambda_{eff}}{\lambda}\quad\mbox{with}\quad \lambda_{eff}=\frac{Q L}{A \Delta T} \mbox{.}
\end{equation}
The applied heat current is given by $Q$, the heat conductivity of the fluid is $\lambda$, and $A$ is the cross sectional area of the cell.
Significant progress has been made in understanding how heat is transported by the flow and thus, how the Nusselt number depends on \Ra\ and \Pra\ \cite[]{AGL09}. Nonetheless there remain unanswered questions. 

In turbulent RBC heat is transported by fluid motion which in part is driven by plumes that detach from thermal boundary layers at the bottom or top plate and rise or sink due to their buoyancy \cite[]{Ka01,SQTX03,XLX04,FA04}. These plumes are carried by, and by virtue of their buoyancy in turn drive, a larger coherent flow structure known as the large-scale circulation (LSC). 
For cylindrical cells with aspect ratio $\Gamma \equiv D/L = 1$ ($D$ is the diameter) the LSC consists of a  single convection roll and is known as a single-roll state or SRS.  Once in a while the SRS collapses but then quickly re-establishes itself. The collapse is known as a cessation \cite[]{BA06a}.   For smaller $\Gamma$ (and \Pra\ near five) two counter-rotating rolls, one on top of the other, were found as well \cite[]{XX08,WA11a} and are  known as a double-roll state or DRS . Then the system  fluctuates randomly  between the DRS and the SRS. The fraction of time $w_{SRS}$ spent in the SRS  increases with increasing \Ra, while the fraction $w_{DRS}$ spent in the DRS decreases. It was found that the SRS transports heat slightly more efficiently than the DRS, but the difference is only about one to two percent depending on \Ra\ \cite[]{WA11a}. Thus the conditional Nusselt number $\Nu_{SRS}$ of the SRS  is larger than $\Nu_{DRS}$ of the DRS.
The  long-time average, designated simply as  \Nu, depends on $w_{SRS}$  and $w_{DRS}$. A bi-modality, where for different measurements at the same \Ra\ two different values of \Nu\ were found \cite[]{RCCH02}, or long transients within a single experiment \cite[]{CRCC04}, cannot be explained by the existence of the SRS and the DRS over the parameter range explored by \cite{WA11a} because they found that  the system freely samples both states over reasonable time intervals.

In the past, experiments were done in which the rotational symmetry of the system was broken by  inclining the cylinder axis at a small angle relative to gravity. 
Studying these slightly anisotropic systems is interesting because many natural convection systems are subject to a broken rotational symmetry ({\em e.g.} convection on hillsides). But in addition one can learn something about the LSC and its shape in containers with a vertical axis.
\cite{CCL96} used a sample with a rectangular cross section and $\Pra \simeq 3$, and tilted their container by 10 degrees in order to fix the orientation of the LSC plane. Within their resolution of a percent or so  they did not find a change in the heat transport. However, they found a significant  reduction of the temperature fluctuations close to the bottom plate. Similar experiments were done by \cite{CCS97} for fluids with $\Pra \simeq 0.025$ in a cylindrical sample with $\Gamma = 1.0$. They applied a tilt of 0.5 degrees in order to create a preferred orientation of the LSC. They also measured the heat flux through the cell for tilt angles up to 4 degrees and found that { '' [...] the heat transfer is unchanged (by no more than 1\% even for much higher tilt of 4 degrees),  [...]''}.
A more comprehensive study of the influence of the tilt angle on the heat transport in $\Gamma=1.0$ cylinders was done by \cite{ABN06} for $\Pra = 4.4$. They found a very small reduction of \Nu\ by about 0.5\% for a tilt angle of 10\textdegree. The changes of the amplitude, and of the dynamics, of the LSC were studied as well and found to be substantial.
Such a  small reduction of the heat transport differs from experimental results reported for cylinders with aspect ratio $\Gamma=0.5$. 
An investigation by \cite{CRCC04} found a decrease of \Nu\ of  up to 5\% for tilt angles as small as $\beta=1.5$\textdegree. 
Another set of measurements by \cite{SXX05} also found a rather large decrease of \Nu, namely by about 2\% for a tilt angle of 2\textdegree.
While these two publications report convection with water and \Pra=4.38, experiments with Helium ($\Pra \simeq 1$) where done by \cite{RGKS10}. They made measurements for two different tilt angles and found that for tilts of 3.6\textdegree\ the heat transport was reduced by about 1.5\% in comparison to that for tilt angels of 1.3\textdegree. 
The above results suggest that there is a difference  between the influence of a tilt on $\Gamma=1$ cells (almost no reduction of \Nu) and the effect seen in $\Gamma=0.5$ experiments (reduction of a few per cent even for very small tilt angles). However, our present study for $\Gamma = 0.50$ does not confirm these results, and instead finds a very small increase of the heat current.

Starting with the Navier-Stokes equation and making physically reasonable approximations, \cite{BA08a} developed a model consisting of two coupled stochastic ordinary differential equations that describes the dynamics of the angular orientation and the amplitude of the LSC in containers with aspect ratio $\Gamma=1$. This model was extended later  to systems with broken rotational symmetry, including samples tilted relative to gravity \cite[]{BA08b}. Generally good agreement with experimental observations was found.
Whereas in principle this model should apply to the SRS independent of $\Gamma$ (albeit with coefficients that depend on $\Gamma$), an extension of the model to the DRS would require a larger number of coupled equations and to our knowledge has not yet been attempted.  

In this paper, we present a comprehensive experimental investigation of the influence of a tilt on the heat transport and the flow structure in a cylindrical  RBC sample with aspect ratio $\Gamma=0.50$. In addition to measurements of \Nu, we analyze the LSC structure and examine the transitions between the SRS and the DRS. One of our findings is that the time $w_{SRS}$ spent in the SRS increases with tilt angle while $w_{DRS}$ decreases. This result is not surprising because in the SRS gravity can favor both up-flow under the tilted top plate and down-flow under the tilted bottom flow, whereas for the DRS only the flow under one of the plates can be favored at the same time \cite[]{CRCC04}. For angles larger than about 6\textdegree\ the DRS no longer exists and $w_{SRS} \simeq 1$.  We found that the increase of $w_{SRS}$ is accompanied by a small increase of \Nu. This too is to be expected because $\Nu_{SRS}$ is larger than $\Nu_{DRS}$ \cite[]{WA11a}, but differs from results of previous investigations \cite[]{CRCC04,SXX05,RGKS10}. Whenever possible, we compare our \Nu\ results with those of  \cite{ABN06} ($\Gamma=1.0$), \cite{BA08b} ($\Gamma=1.0$), and \cite{CRCC04} ($\Gamma=0.5$). 

In addition to the \Nu\ measurements, we report on the LSC amplitude, on a Fourier decomposition which gives its mode structure, and on the probability distribution of the LSC orientation.

\section{Experimental setup and methods}

We used the medium convection apparatus (MCA) described by \cite{ZA10} and \cite{WA11a}. 
Water at an average temperature $T_{m}=(T_b+T_t)/2=40.00$\,\textcelsius\ (\Pra=4.38) was confined between two copper plates, separated by a distance $L=495$\,mm, and a cylindrical Plexiglas sidewall of inner diameter $D=247.5$\,mm, resulting in an aspect ratio $\Gamma=0.500$. The sidewall thickness was 6 mm.
Three sets, each consisting of eight thermistors distributed uniformly in the azinuthal direction, were located at the vertical positions $z=L/4$, L/2 and 3L/4 above the bottom plate.  
All other details were as described by \cite{ZA10} and \cite{WA11a}.

We did experiments for $\Ra=1.8\times 10^{10}$ ($\Delta T = 3.98$ K) and $7.2\times 10^{10}$ ($\Delta T = 15.88$ K) and tilt angles over the range $0\leq\beta  \leq 0.12$\,rad. 
In a typical run we held $\Delta T$ constant and measured the temperatures at all of the thermistors every 3.2\,sec for about a day. After that, the tilt angle was changed. In addition, for a few measurements we took data for very long times (up to 140 hours) to ensure that the statistics does not change for longer measurement times.
Data that were taken within the first five hours of the experimental run were not considered in the data analysis to avoid transient behavior of the system. 

We used the 24 sidewall thermistors to measure the orientation and dynamics of the LSC \cite[]{BNA05}. 
To lowest order, the LSC is a flow where warm fluid rises at one side of the cell and cold fluid sinks at the opposite. Thus, one can detect the azimuthal orientation and the strength of the LSC by fitting a sinusoidal curve of the form
\begin{equation}
T_f=T_{w,k}+\delta_k \cos \left( \frac{\textrm{i}\pi}{4}-\theta_k \right)\mbox{,}
\label{eq:cos}
\end{equation} 
to all eight thermistors at one specific height. Here the index "i" stands for the azimuthal location of the thermistors and takes values $i=0\dots 7$. The index "k" denotes the vertical location of the thermistor and will in the following have letters "b" (z=L/4), "m" (z=L/2) and "t" (z=3L/4). In this way, the amplitude $\delta_k$ is a measure for the strength of the LSC at height level "k" whereas the orientation of the flow is given by the phase $\theta_k$.

\section{Results}

\subsection{Nusselt number measurements}

Nusselt-number measurements in the absence of a tilt ($\beta = 0$) taken with this apparatus were published before (see Fig. 3 of  \cite{WA11a}) and agree with previous measurements and the predictions of the Grossmann-Lohse model \cite[]{GL01} to within a percent or so. For the two Rayleigh numbers investigated in the present paper, \Nu\ in the absence of a tilt had the values $\Nu(0)=160.4$ for $\Ra=1.8\times 10^{10}$ and $\Nu(0)=246.4$ for $\Ra=7.2\times 10^{10}$. 
Here we focus on the effect of a tilt and normalise the measured Nusselt numbers by $\Nu(0)$. In order to improve the accuracy of the normalization, we use the reduced Nusselt numbers $\Nu_{red}(\beta)=\Nu(\beta)/\Ra^{0.3}$ rather than \Nu. In this way we account for small variations of \Ra\ between different experimental runs. 

\begin{figure}
\begin{center}
\includegraphics[width=0.95\textwidth]{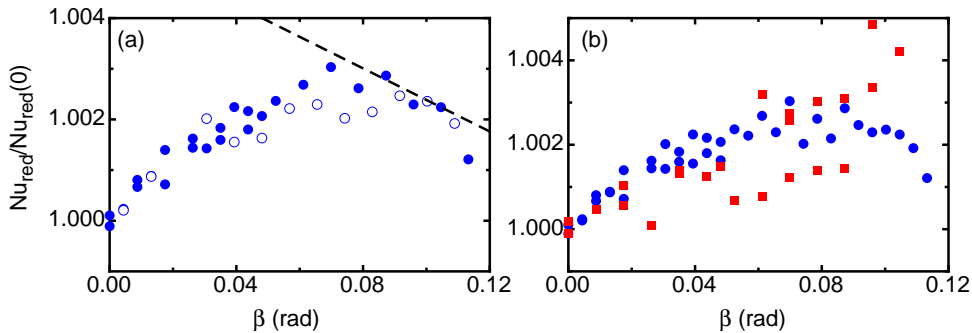}
\caption{The reduced Nusselt number $\Nu_{red}/\Nu_{red}(0)$, with $\Nu_{red} = \Nu/\Ra^{0.300}$,  as a function of the tilt angle $\beta$. (a): Data for $\Ra=7.2\times 10^{10}$ and for increasing tilt (solid symbols) and decreasing tilt (open symbols). The dashed line has a slope equal to the one measured for $\Gamma = 1.00$ by \cite{ABN06}. (b): Data for both $\Ra=1.8\times 10^{10}$ (squares, red online) and $\Ra=7.2\times 10^{10}$ (bullets, blue online). Data points in (b) include data for increasing and decreasing tilt angles.}
\label{fig:Nu-beta}
\end{center}
\end{figure}

Figure \ref{fig:Nu-beta}a shows the normalised Nusselt number $\Nu_{red}(\beta)/\Nu_{red}(0)$ as a function of the tilt angle $\beta$ for $\Ra=7.2\times 10^{10}$. In order to check for hysteretic (history dependent) effects, we plot measurement points that were taken while $\beta$ was increased as solid symbols, whereas measurements with decreasing $\beta$ are shown as open symbols.
The data for increasing and decreasing tilt agree within their range of scatter and thus no hysteretic effect is visible here.

The behavior of $\Nu(\beta)/\Nu(0)$ differs significantly from measurements reported by \cite{CRCC04}, \cite{SXX05} and \cite{RGKS10}.
We observe a very small but well resolved increase of the heat transport with increasing tilt angle.
The heat transport reaches a maximum near $\beta\approx 0.07 $ rad and decreases for larger $\beta$. Over the investigated range of $\beta < 0.113$\,rad \Nu\  is always larger than it is in the horizontal case  $\beta = 0$. 
However, the increase is very small, with a maximum \Nu\ increase of only about 0.3\%.

Similar behavior, but with larger scatter of the data, is also observed for $\Ra=1.8\times 10^{10}$ as marked by squares (red online) in figure \ref{fig:Nu-beta}b. The larger scatter is due to the fact that the smaller value of \Ra\ corresponds to a smaller $\Delta T$ where the same temperature resolution leads to a larger relative scatter.  
Due to the scatter of the small-\Ra\ data it can not be concluded whether or not there exists a maximum in the range $0<\beta<0.12$. Nevertheless an increase of the heat transport with increasing tilt angles is clearly visible in the data.

The observation of a small increase of heat transport with increasing tilt seems unexpected and is in contrast to previous observation \cite[]{CRCC04,SXX05,RGKS10}. However, we shall show that the reason for this increase can be found in a stabilization of the SRS relative to the DRS.
Since the SRS transports heat more efficiently than the DRS by about one to two percent depending on \Ra, a small net increase of \Nu\ can be explained.
For this purpose we analyze the side-wall measurements in the following subsection.

\subsection{Flow-mode transition}
\label{sec:SRS-DRS} 

As mentioned in the introduction, studies on RBC in cylinders with $\Gamma=0.50$ \cite[]{XX08,WA11a} have shown that in these geometries and for $\Pra \simeq 5$, the large-scale circulation can exist in form of a single-roll state (SRS) or a double-roll state (DRS). 
The latter consists of two counter-rotating rolls, one on top of the other. 
The flow switches randomly between the two states. The fraction of time that the system spends in the DRS depends on \Ra, but is in general small in comparison to the fraction of time the system exhibits a SRS. 
It was shown (see Fig. 19 of \cite{XX08} and Fig. 14 of \cite{WA11a}) that the heat transport is less efficient when the DRS is present and thus the Nusselt number is reduced slightly in comparison to the times when the system exhibits the SRS. 

In a slightly inclined convection cell one would expect the SRS to be favored by the system, since in the DRS the buoyancy adjacent to either the inclined cold (top) or the inclined warm (bottom) plate acts against the direction of the fluid flow, depending on the orientation of both rolls \cite[]{CRCC04}.

To detect the state of the system, we look at the amplitudes $\delta_k$ and phases $\theta_k$ at the three different levels $z = L/4, L/2$ and $3L/4$. As described by \cite{WA11a}, we define the system to be in the SRS when $|\theta_t-\theta_b|<60$\textdegree\ and demand that the amplitudes at all levels k="t","m" and "b" be larger than 15\% of their average values ($\delta_k>0.15\langle \delta_k\rangle $). On the other hand, we say that the system is in the DRS when $|\theta_t-\theta_b|>120$\textdegree\ and $\delta_{t,b}>0.15\langle \delta_{t,b}\rangle$. 
When neither the conditions for a SRS nor those for the DRS are fulfilled, we call this a {\em transition state} (TS). The largest contribution to the TS comes from states for which  $60 < |\theta_t-\theta_b| < 120$\textdegree.

\begin{figure}
\begin{center}
\includegraphics[width=\textwidth]{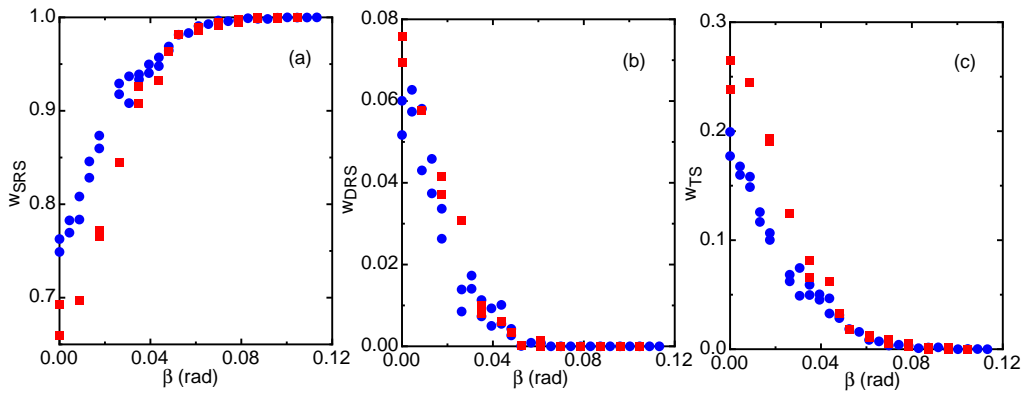}
\caption{(a): The fraction of time $w_{SRS}$ that the system exists in the single-roll state. (b): The fraction of time $w_{DRS}$ that the system exists in the double-roll state. (c):  The fraction of time  $w_{TS}$ that the system exists in the transition state. The data are for $\Ra=1.8\times 10^{10}$ (red squares) and $\Ra=7.2\times 10^{10}$ (blue bullets).}
\label{fig:SRS-DRS}
\end{center}
\end{figure}

Figure \ref{fig:SRS-DRS} shows the fraction of time that the system spends in the SRS (a), in the DRS (b), and in the TS (c) for the two investigated Rayleigh numbers. The result is clear and as expected. With increasing tilt angle, the system spends more time in the SRS and less in the DRS or TS. For angles larger than $\beta\approx 0.06$\,rad the system consists essentially at all times of a single well defined roll, where warm fluid flows along the bottom plate in the uphill direction and cold fluid flows along the top plate in the downhill direction.

\begin{figure}
\begin{center}
\includegraphics[width=0.6\textwidth]{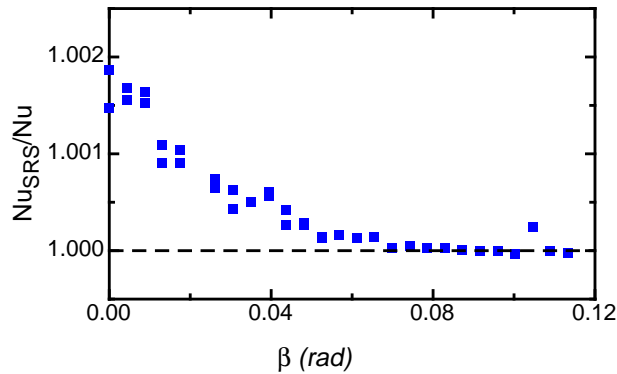}
\caption{The conditional Nusselt number $\Nu_{SRS}$ computed only from data taken while the system was in the SRS. The results were normalised by \Nu\ and are shown as a function of the tilt angle. They are for $\Ra=7.2\times 10^{10}$. The dashed horizontal line marks $\Nu_{SRS}/\Nu=1$.}
\label{fig:Nu_SRS}
\end{center}
\end{figure}

\cite{WA11a} showed that the heat is transported more effectively when the system is in the SRS.
Therefore, we believe that the preference of the inclined system for the SRS explains the increase of \Nu\ with $\beta$. We note, that the location of the maximum of $\Nu(\beta)/\Nu(0)$ at $\beta\approx 0.07$\,rad roughly coincides with the angle above which the system exhibits the SRS during nearly the whole measurement time. Thus, a further increase of $\beta$ cannot result in a further increase of \Nu\ and from there on $\Nu$ decreases gradually.  
To illustrate this fact in more detail, we calculate the conditional Nusselt number $\Nu_{SRS}$  that takes only the time intervals into consideration when the system is in the SRS. This quantity, divided by \Nu, is plotted as a function of $\beta$ in Fig. \ref{fig:Nu_SRS}.
The plot shows two things. First, for no or small tilt angles $\Nu_{SRS}$ is larger than \Nu, stating again that the heat is transported more efficient in the SRS than in the DRS or the TS. Second, for increasing tilt angles $\Nu_{SRS}$ becomes more nearly equal to \Nu\ since the SRS becomes the dominant flow state of the system.

For cylindrical samples with $\Gamma=1.0$ the sample is always in the SRS. Consistent with this and the phenomenon described above, measurements of \Nu\ did not reveal any increase with $\beta$. Instead they showed a very gradual decrease with increasing $\beta$, corresponding to
$\Nu(\beta) = \Nu(0)(1+ a_{\Nu} \beta)$ with $a_{\Nu} = -0.031$  \cite[]{ABN06}.  In Fig.~\ref{fig:Nu-beta} a we show a dashed line which has a slope corresponding to the measured $a_{\Nu}$ for $\Gamma = 1.00$. The present data for $\Gamma = 0.50$ are consistent with a similar decease of $\Nu(\beta)$ in the $\beta$ range beyond the maximum where the initial increase due to a diminished presence of the DRS no longer occurs. However, the data do not extend to sufficiently large $\beta$ to yield quantitative information on this issue.  

In order to compare our results with those of others, we  briefly consider low-Prandtl-number convection in cylinders with $\Gamma=0.5$. On the basis of measurements by \cite{ABFH09}, \cite{WA11a} showed that, for $\Pra=0.67$, $\Ra=1.0\times 10^{11}$, and $\Gamma = 0.50$, the system is in a SRS almost all the time. They found no evidence for a DRS. Thus, we expect no increase of \Nu\ with $\beta$ for this case, and by analogy to the $\Gamma = 1.00$ measurements a slight decrease of \Nu\  with $\beta$ might be expected.  Consequently, the reduction of the heat transport by about 2\% with a change of $\beta$ from 0.023 to 0.063 rad in the \Ra\ range from $6\times 10^{10}$ to $2\times 10^{11}$ reported by \cite{RGKS10} is in qualitative agreement with our measurements. However, the reduction of $\Nu$ by 2\%, corresponding to a slope $a_{\Nu} \simeq -0.5$, differs by about a factor of 16 from  the measured value \cite[]{ABN06} for $\Pra = 4.3$ and $\Gamma = 1.00$. Thus, in order to reconcile the data of  \cite{RGKS10} quantitatively with our measurements and those of \cite{ABN06}, a very strong \Pra\  dependence of $a_{\Nu}$ would have to be invoked. In view of recent numerical calculations \cite[]{BMS10,PSL11}, which revealed that the nature of the LSC has a larger influence on \Nu\ as \Pra\ decreases to or below one, such a strong dependence can not be ruled out.

The reason for the difference between our data and the heat-transport measurements as a function of $\beta$ by \cite{CRCC04} and \cite{SXX05} (with $\Pra \simeq 5$ similar to our case) remains unexplained.

\section{Properties of the LSC}

\subsection{The average amplitude of the LSC}
\label{sec:delta}

\begin{figure}
\begin{center}
\includegraphics[width=0.9\textwidth]{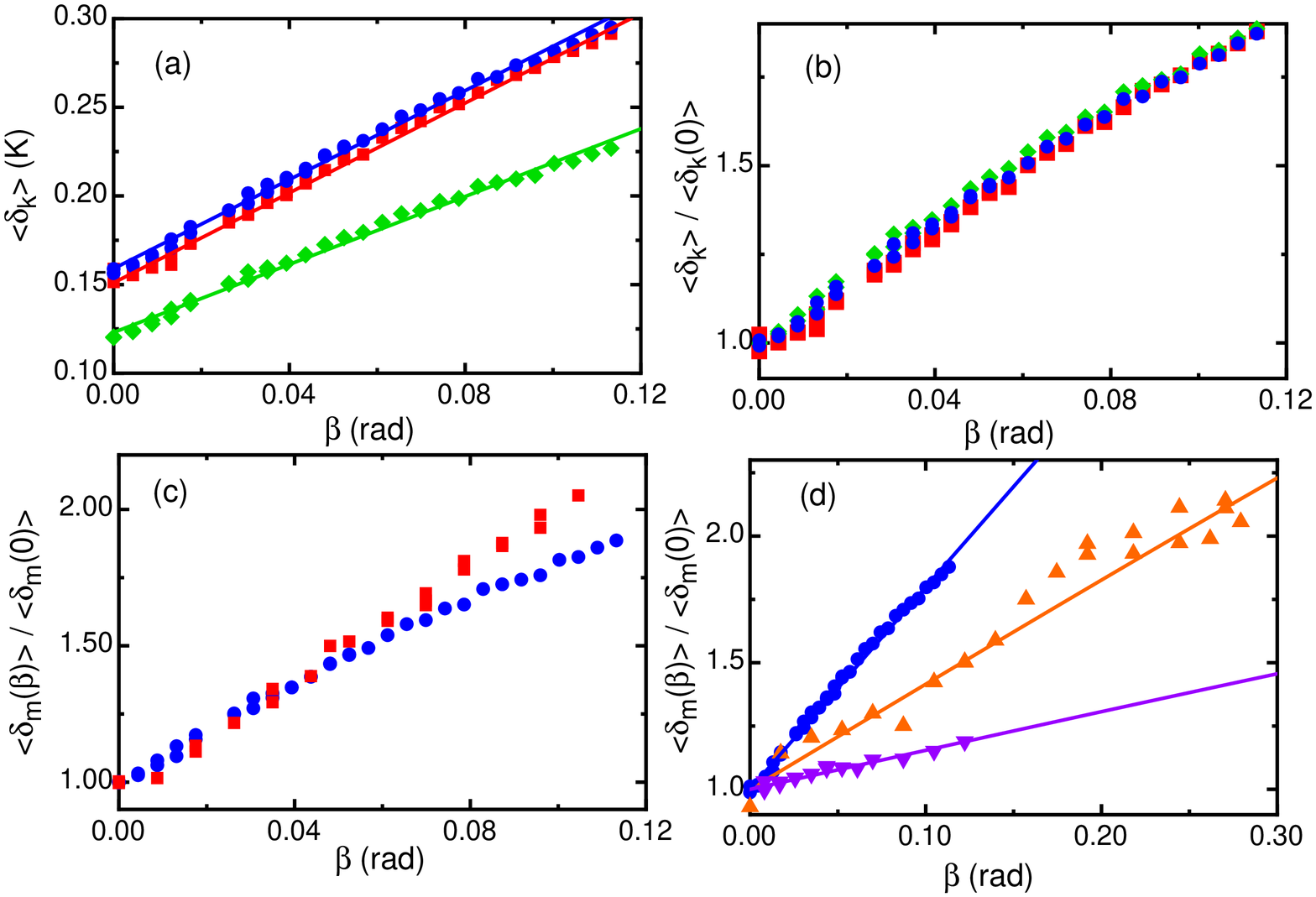}
\caption{The time-averaged amplitude $\langle \delta_k \rangle$ of the azimuthal temperature variation along the side wall as a function of the tilt angle $\beta$. (a): The amplitude for the top ($\langle\delta_t\rangle$, bullets, blue online), the middle ($\langle\delta_m\rangle$, diamonds, green online) and the bottom ($\langle\delta_b\rangle$, squares, red online) thermistor row. The data were taken at $\Ra=7.2\times 10^{10}$. (b): The same data as in (a) but normalised by their values at $\beta=0$. (c): Comparison of the normalised amplitude at midheight $\langle\delta_m(\beta)\rangle/\langle\delta(0)\rangle$ for $\Ra=1.8\times 10^{10}$ (squares, red online) and $\Ra=7.2\times 10^{10}$ (bullets, blue online). (d): Comparison between $\Gamma=0.50$ data (bullets, blue online, $\Ra=7.2\times 10^{10}$) and data for $\Gamma=1.0$ [up-pointing triangles, yellow online, $\Ra=2.8\times 10^9$, from \cite{BA08b}; and down-pointing triangles, purple online, $\Ra = 9.4\times 10^{10}$, from \cite{ABN06}]. 
The solid lines in (a) and (d) are fits of Eq. \ref{eq:para} to the data. The coefficients derived from those fits are shown in Table \ref{tab:para}.  }
\label{fig:delta}
\end{center}
\end{figure}

Here we investigate the time-averaged strength of the LSC as a function of the tilt angle $\beta$. 
Figure \ref{fig:delta}a shows the dependence of $\langle \delta_k\rangle$ on $\beta$ for $\Ra = 7.2\times 10^{10}$. For no tilt, $\langle\delta_m\rangle$ is significantly smaller than $\langle \delta_t\rangle$ and $\langle \delta_b\rangle$. This finding is in contrast to results for $\Gamma=1.00$, where $\langle\delta_m\rangle$ was larger \cite[]{BA07_EPL}, but in agreement with previous results for $\Gamma=0.50$ \cite[]{WA11a}. 
With increasing tilt angles all three $\langle\delta_k\rangle$ increase.
Such an increase was also found for $\Gamma=1$, where however only the temperature at the horizontal mid-plane ($\langle\delta_m\rangle$) was investigated as a function of $\beta$ \cite[]{ABN06}. 
The increase of the $\langle\delta_k\rangle$ is smooth and does not show any obvious change at or beyond $\beta\approx 0.06$ where the DRS and TS cease to exist. 

The increase of $\langle\delta_k\rangle$ with $\beta$ is not linear. The slope $\partial \langle\delta_k\rangle/\partial \beta$ decreases with increasing $\beta$. This is also in agreement with the results for $\Gamma=1$ \cite[]{ABN06}. 
Based on the model for the LSC by \cite{BA08b}, we fit the sinusoidal function
 
\begin{equation}
\label{eq:para}
\langle\delta_k\rangle = \langle\delta_k(0)\rangle\cdot[1+a \sin (\beta)]
\end{equation}

\noindent to the data for a quantitative comparison. The corresponding fits are shown as solid lines in figure \ref{fig:delta}a. The coefficients $\langle\delta_k(0)\rangle$ and $a$  are listed in table \ref{tab:para}.
The coefficient $a$ does not depend significantly on the level.
The amplitudes, normalised by their values $\langle\delta_k(0)\rangle$ for $\beta = 0$, are plotted in figure \ref{fig:delta}b.
One sees that the {\em relative} increase of $\langle\delta_k\rangle$ is the same for all levels.
We shall show below in Sec. \ref{sec:delta_dist} that this is the case also for the probability distributions of the $\langle\delta_k\rangle$. 

\begin{table}
\begin{center}
\begin{tabular}{ccccccc}
\hline
$\Gamma$  & $\Delta T$ (K) & \Ra & level & $\langle\delta_k(0)\rangle (K)$ & $a$ & Reference \\ 
\hline
0.50 & 3.98  & $1.8\times 10^{10}$ & top		& 0.050  & $ 11.3 \pm 0.3$ & this work \\
0.50 & 3.98  & $1.8\times 10^{10}$ & middle	&  0.042 & $ 11.1\pm 0.3$ & this work  \\
0.50 & 3.98  & $1.8\times 10^{10}$ & bottom	&  0.048 & $ 11.6\pm 0.5$ & this work  \\
0.50 & 15.88  & $7.2\times 10^{10}$ & top	& 0.159 & $7.9\pm 0.1$ & this work \\
0.50 & 15.88  & $7.2\times 10^{10}$ & middle	& 0.123 & $7.8\pm 0.2$ & this work  \\
0.50 & 15.88  & $7.2\times 10^{10}$ & bottom	& 0.151 & $8.4\pm 0.2$ & this work  \\
1.00 &  4.96   & $2.8\times 10^{9}  $ & middle 	& 0.023 &  $4.2\pm 0.3$ & \cite{BA08b} \\
1.00 & 19.63  & $9.4\times 10^{10}$ & middle 	& 0.165 & $1.5\pm 0.1$ & \cite{ABN06} \\
\end{tabular}
\caption{Coefficients from fits of Eq.~\ref{eq:para} to data for $\langle\delta_k(\beta)\rangle$. All results are for $\Pra = 4.38$.}
\end{center}
\label{tab:para}
\end{table}

To compare the strength of the LSC for both \Ra\ values that we investigated, we plot in Fig.~\ref{fig:delta}c the amplitude at mid-height $\langle\delta_m(\beta)\rangle/\langle\delta_m(0)\rangle$ for $\Ra=1.8\times 10^{10}$ (squares, red online) and for $\Ra=7.2 \times 10^{10}$ (bullets, blue online). 
One sees that for the larger \Ra\ the curve starts to bend earlier.
This might suggest that a simple sinusoidal fit is not sufficient to characterise $\langle\delta_k(\beta)\rangle$, and that higher-order terms might be needed. Nonetheless, for the present purpose we shall retain the use of Eq.~\ref{eq:para}.

In Fig. \ref{fig:delta}d and Table~\ref{tab:para} we compare our measurements of $\langle\delta_m\rangle$ ($\Gamma=0.50$) with measurements for $\Gamma=1.00$ at $\Ra=2.8\times 10^9$ \cite[]{BA08b} and at  $\Ra = 9.4\times 10^{10}$ \cite[]{ABN06}. One sees that at nearly the same \Ra\  the increase of $\langle\delta_m\rangle$  is stronger (the coefficient $a$ is larger) for $\Gamma=0.50$ than it is for $\Gamma=1.00$. The data also suggest a significant decrease of $a$ as \Ra\ increases at constant $\Gamma$.   

\begin{figure}
\begin{center}
\includegraphics[width=0.6\textwidth]{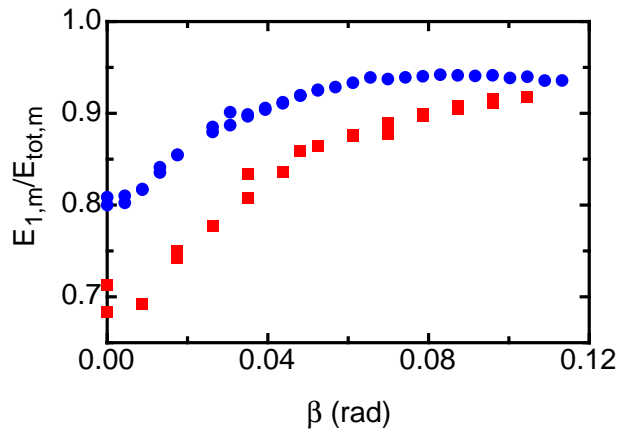}
\caption{The ratio of the energy in the first Fourier mode $E_{1,m}$ to the total energy of all four modes $E_{tot}$ for $\Ra=1.8\times 10^{10}$ (squares, red online) and for $\Ra=7.2\times 10^{10}$ (bullets, blue online). The measurements here are for the middle thermistor row. Measurements for the top and the bottom give nearly the same result.}
\label{fig:LSC_Modes}
\end{center}
\end{figure}

\subsection{Higher-order modes}

In the previous section $\langle\delta_k\rangle$ was obtained from a fit of Eq.~\ref{eq:cos} to the side-wall temperature-measurements, and thus it represents the amplitude of the first Fourier mode of the azimuthal temperature variation.
To gain information about the shape of the LSC it is useful also to compare the energy $E_{1,k} = \delta_k^2$ of this lowest mode with that of the higher modes \cite[]{SCL11,WA11c}.
Since the behavior is quite similar for all three thermistor rows, we discuss only the analysis for the middle row.
Figure \ref{fig:LSC_Modes} shows the ratio between the energy $E_{1,m}$ of the first mode to the total energy $E_{tot,m}$ of all four accessible modes (we have eight thermistors at one height and thus have access to four Fourier modes). 
The data show an increase of $E_{1,m}/E_{tot,m}$ with increasing tilt angle. One sees that higher-order modes are suppressed by increasing the tilt.
We also see that $E_{1,m}/E_{tot,m}$ is smaller for $\Ra=1.8\times 10^{10}$ than it is for $\Ra = 7.2\times 10^{10}$. However, as $\beta$ increases,  $E_{1,m}/E_{tot,m}$ reaches similar values, larger than 0.9, for both \Ra.

\subsection{Events}
\label{sec:events}

An event is defined as a drop of the amplitude $\delta_k$ below 15\% of its time-averaged value $\langle \delta_k\rangle$.
In this section we report on the time-averaged event frequency $\omega_k^e$ as a function of the tilt angle. 
The cause of events may vary and depends on $\Gamma$.
While for $\Gamma=1.0$ most of the events are cessations [almost no flow-mode transitions are found  \cite[]{XX08}], most of the events for $\Gamma=0.50$ are caused by flow-mode transitions [cessations are very rare \cite[]{WA11a}].

\begin{figure}
\begin{center}
\includegraphics[width=0.95\textwidth]{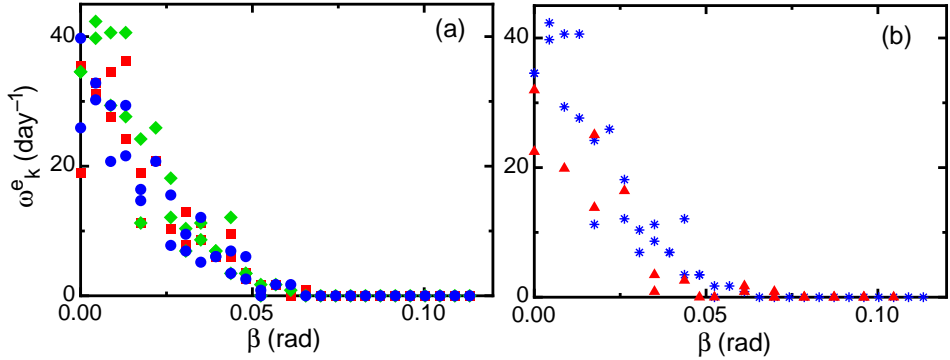}
\caption{The time-averaged frequencies $\omega_k^e$ at which the amplitudes $\delta_k$ drop below 15\% of their average values. (a): Frequency of events for the top (bullets, blue online), the middle (diamonds, green online) and the bottom (squares, red online) thermistor row. Measurements were taken at $\Ra=7.2\times 10^{10}$. (b): Frequency of events for the middle thermistor row for $\Ra=1.8\times 10^{10}$ (triangles, red online) and $\Ra=7.2\times 10^{10}$ (stars, blue online).}
\label{fig:events}
\end{center}
\end{figure}

Figure \ref{fig:events}a shows the event rate for the three different heights as a function of the tilt angle $\beta$.
The data points show considerable scatter because the number of events that occurred during the measurement times of about a day for each point was small. For $\beta=0$ the event rate is comparable to previous measurements \cite[]{WA11a}.
No significant difference can be observed between the event rate of the top, the middle, and the bottom thermistor row.
This can be understood since these events are mostly flow-mode transitions at which a smaller roll appears at the top or the bottom, grows in size and replaces the original roll.
During such transitions a dip (an event) can be observed in all of the three amplitudes (note however, that this is also true for cessations).

The decrease of $\omega_k^e$ with increasing tilt angle, and the observation that for $\beta > 0.06$ no events occurred, 
is consistent with the results for the fraction of time $w_{SRS}$ that the system spent in the SRS as reported above in Fig.~\ref{fig:SRS-DRS}a.
As the SRS becomes more dominant and the DRS becomes rare, flow-mode transitions ({\it i.e.} transitions between these two states) become less frequent. When, for $\beta > 0.06$ only the SRS is found, no transitions can occur.  
The finding of a reduction of events with increasing tilt is in accordance to the results for $\Gamma=1.00$ \cite[]{BA08b}.
There however, the event rate was already significantly smaller in the horizontal case (1.7 per day) and was mainly caused by cessations.

In Fig. \ref{fig:events}b we compare the event rates of the middle thermistor row $\omega_m^e$ for experiments with $\Ra=1.8\times 10^{10}$ and $\Ra=7.2\times 10^{10}$.
For $\beta < 0.05$ the data suggest a slightly smaller event rate for the smaller \Ra, but due to the scatter a definitive conclusion can not be drawn.

\subsection{The probability distribution of $\delta_k$}
\label{sec:delta_dist}

\begin{figure}
\begin{center}
\includegraphics[width=0.95\textwidth]{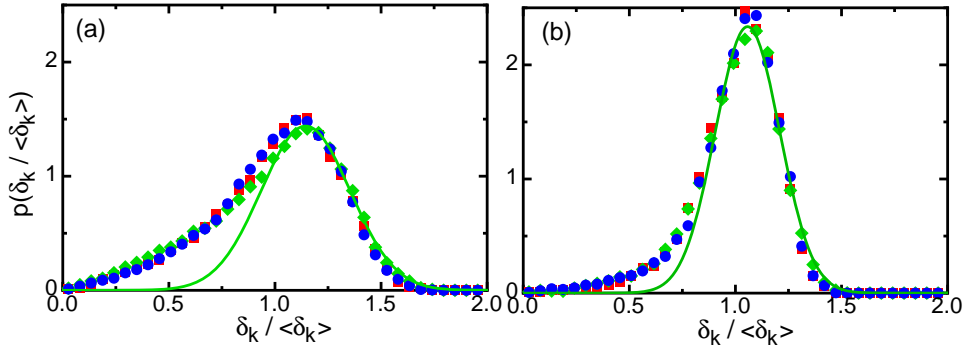}
\caption{The probability-density functions of the normalised LSC strength $\delta_k/\langle \delta_k \rangle$ for the top (bullets, blue online), the middle (diamonds, green online) and the bottom (squares, red online) thermistor row. Shown are data for (a) $\beta=0$ and (b) $\beta=0.035$\,rad. The solid lines (green online) are Gaussian fits to the right side of the peak of $\delta_m/\langle \delta_m \rangle$. The experiments were done at $\Ra=7.2\times 10^{10}$.}
\label{fig:delta_PDF}
\end{center}
\end{figure}

We showed previously \cite[]{WA11a} that the probability-density functions (PDFs) $p(\delta_k)$ of the amplitudes $\delta_k$ at different levels $k$ collapse when the $\delta_k$ are scaled by their corresponding time averaged values $\langle \delta_k\rangle$. 
This is well illustrated by figure \ref{fig:delta_PDF}a where for the horizontal case ($\beta = 0$) the PDF's of the normalised amplitudes are plotted. The graph looks very similar to the one shown by \cite{WA11a}. The PDF's are asymmetric. At $\delta_k/\langle \delta_k\rangle=0$ they start with a finite slope that increases until an inflection point is reached. From there on $p(\delta_k/\langle \delta_k\rangle)$ continues to grow until it reaches a maximum. The maximum and the right tail of the PDF can be described fairly well by a Gaussian distribution as indicated by the solid lines in figure \ref{fig:delta_PDF}. All this is consistent with a description of $\delta_k$ as a stochastically driven amplitude diffusing in an asymmetric potential which was derived by \cite{BA08a} and extended recently by \cite{AAG11}, albeit with parameters for $\Gamma = 1.00$.  

Figure \ref{fig:delta_PDF}b shows data similar to those in (a), but here the cell was tilted by $\beta=0.035$\,rad. The effect on $p(\delta_k/\langle \delta_k\rangle)$ is clearly visible. The peak becomes significantly narrower (and thus higher) than it is in the horizontal case. Since also the slope at $\delta_k/\langle \delta_k \rangle=0$ becomes smaller, the peak becomes more nearly symmetric. This result is in accordance with the reduction of flow-mode transitions and the stabilization of the SRS since flow-mode transitions and other events (see section \ref{sec:events}) result in a temporary decrease of $\delta_k$, and thus in a skewed distribution.

\begin{figure}
\begin{center}
\includegraphics[width=0.95\textwidth]{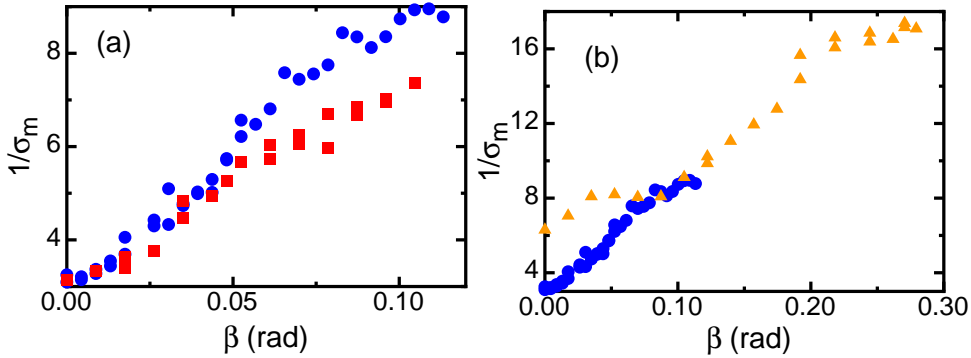}
\caption{Inverse of the standard deviation $\sigma_m$ of the amplitude distribution as a function of the tilt angle $\beta$. (a): Comparison between $\Ra=1.8\times 10^{10}$ (squares, red online) and $\Ra=7.2\times 10^{10}$ (bullets, blue online).
(b): Comparison between $\Gamma=0.50$ (bullets, blue online, $\Ra=7.2\times 10^{10}$) and $\Gamma=1.00$ (triangles, yellow online, $\Ra=2.8\times 10^8$, from \cite{BA08b}).}
\label{fig:sigma_delta}
\end{center}
\end{figure}

For a quantitative analysis, we calculate the standard deviation
\begin{equation}
\sigma_k \equiv \langle (\delta_k/\langle \delta_k\rangle - 1)^2\rangle^{1/2}
\end{equation}
 of the normalised amplitude for the middle thermistor row $k = m$. We note that a similar calculation for the other thermistor rows gives very similar results, as indicated by figure \ref{fig:delta_PDF}.
In Fig.~\ref{fig:sigma_delta} we show the inverse  $1/\sigma_m$ as a measure of the stability and the coherence of the LSC.
It is interesting that in the horizontal case $1/\sigma_m$ is the same for both investigated \Ra\ values.
However, when the sample is tilted, $1/\sigma_m$ increases somewhat faster for the larger \Ra.

In Fig. \ref{fig:sigma_delta}b  we compare our data ($\Gamma=0.50$) with those for $\Gamma=1.00$ of \cite{BA08b}. Note, that there is a difference of almost three orders of magnitude in \Ra. 
It is not a surprise to see that in the horizontal case the flow is more stable (larger $1/\sigma$) for $\Gamma=1.00$ than it is for $\Gamma=0.50$. This was already found and discussed in previous publications \cite[]{XX08,XX08a,WA11a}. 
However, it is surprising that for $\beta>0.8$ the $\Gamma=1.00$ data show the same slope as the $\Gamma=0.50$ data did in the $\beta$ range that we investigated. 
This seems remarkable, especially since the  \Ra\ values at which the two data sets were taken differ by two and a half orders of magnitude.     

\subsection{The orientation of the LSC}

By inclining the cell from its horizontal position, the rotational symmetry is broken and the azimuthal orientation $\theta_k$ of the LSC plane is no longer randomly distributed but tends to align in the direction of the tilt. 
This process can be investigated by looking at the probability-density function of $\theta_k$ at the different heights. 

\subsubsection{The horizontal case $\beta = 0$}

\begin{figure}
\begin{center}
\includegraphics[width=0.95\textwidth]{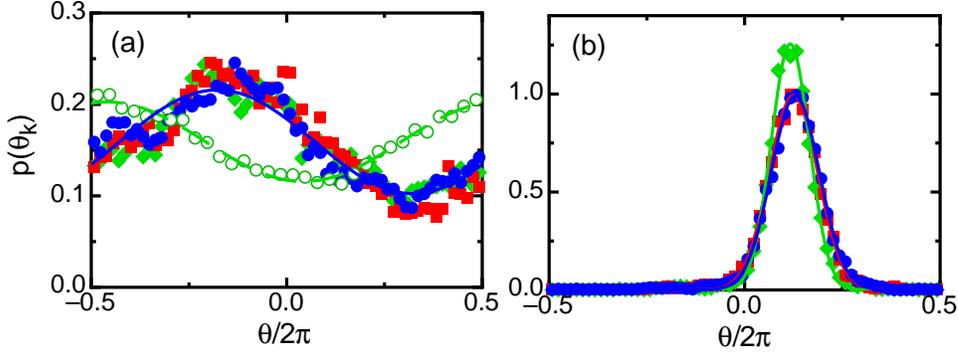}
\caption{Probability density functions of the orientations $\theta_k$ of the LSC plane for $\Ra=7.2\times 10^{10}$ at tilt angles of  (a) $\beta = 0.000$ and (b) $\beta = 0.035$ rad. The data are for the top (bullets, blue online), the middle (diamonds, green online) and the bottom (squares, red online) thermistor row.  The open green circles in (a) show $p(\theta_m)$ from a previous experiment with $\Ra=9.0\times 10^{10}$ and $\beta = 0.000$ \cite[]{WA11a}. The lines in (a) are fits of Eq.~\ref{eq:sinPDF} to the data, and the lines in (b) are fits of Eq. \ref{eq:Gauss_theta} to the data.}
\label{fig:theta_PDF_example}
\end{center}
\end{figure}

For the perfectly horizontal and rotationally invariant system all angles $\theta_k$ should be sampled equally and the (normalised) PDF $p(\theta_k),~ 0 \leq \theta_k \leq 2\pi$, should be a constant equal to $1/(2\pi)$. PDFs for $\theta_k$   at $\Ra = 7.2\times 10^{10}$ are shown in Fig. \ref{fig:theta_PDF_example} for (a) the horizontal case $\beta = 0.000$ and (b) a tilt of $\beta=0.035$\,rad. One sees that already for the horizontal case there is  a small anisotropy. The circulation plane is less likely to be found near $\theta/2\pi\approx 0.3$ and more likely to be found near $\theta/2\pi\approx -0.2$.
The PDF has to be $2\pi$-periodic. It turns out that the simplest function with this property, namely
\begin{equation}
p(\theta_k) = 1/(2\pi) + A_k\cdot \sin(\theta_k - \phi_k)\ ,
\label{eq:sinPDF}
\end{equation}
 fits the data very well. Here the phase $\phi_k$, although it should be equal at all $k$,  is arbitrary since it depends on the arbitrarily selected origin of the azimuthal coordinate system which may vary from one experimental setup to another. A fit of Eq.~\ref{eq:sinPDF} to the data for $k = t$ (the top thermistor row, blue bullets) is shown in Fig.~\ref{fig:theta_PDF_example}a as a solid line (blue online). The fits for $k = m$ (solid diamonds, green online) and $k = b$ (solid squares, red online) are not shown for clarity, but they are equally good. The amplitude $A_k$ is a measure of the extent of the deviations from the expected uniform distribution of the rotationally invariant system. For the example of Fig.~\ref{fig:theta_PDF_example} we found $A_b = 0.068,~ A_m = 0.056$, and $A_t = 0.057$, all with probable errors of 0.003. As expected, there is little if any significant height dependence.

Also shown in Fig.~\ref{fig:theta_PDF_example}a, as open circles (green online),  are results from a previous investigation for $\Ra=9.0\times 10^{10}$ \cite[]{WA11a}). The dashed line (green online) in Fig.~\ref{fig:theta_PDF_example}a is the fit of Eq.~\ref{eq:sinPDF} to those data (in this case the phase $\phi_m$ is different because the arbitrary origin of the azimuthal coordinate was different). That fit gave $A_m = 0.044$, which is somewhat lower than but close to the present result. A third data set, for $\Gamma = 0.50$ and $\Ra=5.7\times 10^{10}$, was reported by \cite{XX08} and suggests similar values of the $A_k$.
 
The inhomogeneity of $p(\theta_k)$ might be due, for instance, to small anisotropies in the experiment, such as for example of the spatial temperature distribution in the top and bottom plates, or to small inhomogeneities of the sidewall \cite[]{BA08b}.
However, in experiments with $\Gamma=1.0$ the anisotropy in $p(\theta_k)$ is significantly stronger and even for the horizontal case the PDF consists of a well defined peak which can be fit well by a Gaussian distribution \cite[]{BA06b,XX08a}. It was shown that the Coriolis force caused by the Earth's rotation leads to a preferred orientation of the LSC \cite[]{BA06b} with a PDF in agreement with the measurement. As pointed out by \cite{XX08a}, according to the model of \cite{BA06b} the influence of the Coriolis force should be even stronger for $\Gamma = 0.50$ than it is for $\Gamma = 1.00$ because the vertical part of the LSC, which causes the preferred orientation,  is longer.
However, the dynamics of the LSC for $\Gamma=0.50$ is more erratic and a SRS is often destroyed by events as documented above in Sec.~\ref{sec:events} and Fig.~\ref{fig:events} ({\em e.g.,} by flow mode transitions) and reappears within short time intervals, often at a different orientations \cite[]{XX08a}. This tends to randomize the orientation, counter-acts the orienting influence of the Coriolis force, and leads to a broader PDF of $\theta_k$.

\subsubsection{The inclined case $\beta > 0$}

The difference between the horizontal case (Fig. \ref{fig:theta_PDF_example}a) and the inclined one (Fig. \ref{fig:theta_PDF_example}b) is clearly visible. At $\beta=0.035$\,rad only orientations close to $\theta/2\pi=0.13$ existed. This shows that the orientation of the LSC is concentrated near one azimuthal position. It also suggests that the single roll is more persistent and is less often destroyed by events (see Fig.~\ref{fig:events}), such as flow-mode transitions or cessations, which would tend to randomize the orientations.

\begin{figure}
\begin{center}
\includegraphics[width=0.95\textwidth]{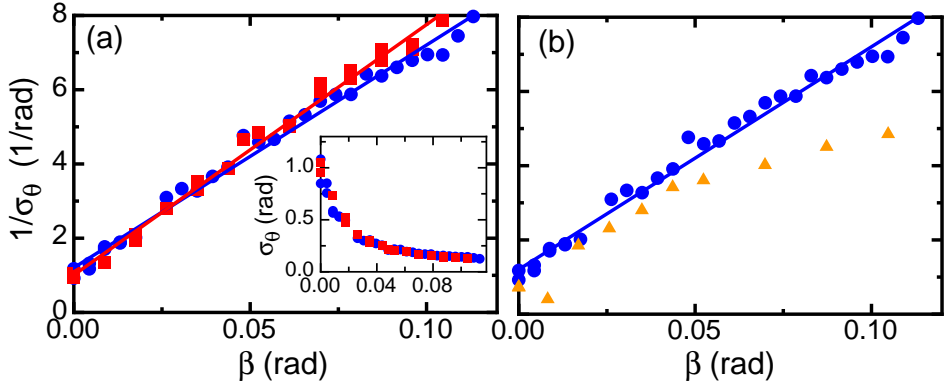}
\caption{The inverse of the standard deviation  $\sigma_\theta$ of the distributions as a function of the tilt angle $\beta$. (a): Comparison between $\Ra=1.8\times 10^{10}$ (squares, red online) and $\Ra=7.2\times 10^{10}$ (bullets, blue online). The solid lines are fits of the straight line $1/\sigma_{\theta} = p + q\cdot \beta$ to the data. The fits yielded the offsets and slopes  $p=1.0 \pm 0.1$, $q= 67\pm 2$ for $\Ra=1.8\times 10^{10}$ and $p = 1.1\pm 0.1$, $q=61\pm 2$ for $\Ra=7.2\times 10^{10}$. The insert shows the same data, but with $\sigma_\theta$ (rather than its inverse) plotted against $\beta$. (b): The bullets (blue online) are for $\Ra=7.2\times 10^{10}$. The triangles (yellow online) are data for $\Gamma=1.00$ at $\Ra=9.43\times 10^{10}$ from \cite{ABN06} (these data were multiplied by $2\pi$ in order to correct a mistake in Fig. 12 of \cite{ABN06}). }
\label{fig:sigma_theta}
\end{center}
\end{figure}

To quantify the narrowing of the PDF, we fit the Gaussian distribution
\begin{equation}\label{eq:Gauss_theta}
p(\theta_m)=A \exp\left(-\frac{(\theta_m-\theta_{m0})^2}{2\sigma_\theta^2}\right)
\end{equation}
to $p(\theta_m$). Although this function is not $2\pi$ periodic (as in principle $p(\theta_k)$ should be), it provides an excellent approximation when the peak is narrow and the tails do not extend too far. Nonetheless, for the fit we  only used points close to the maximum since for small tilt angles one tail of the Gaussian is usually superimposed upon the other due to the $2\pi$-periodicity of $\theta_m$.
The insert in figure \ref{fig:sigma_theta}a shows the decrease of $\sigma_\theta$ with increasing tilt angle.
However, by plotting its inverse $1/\sigma_\theta$ in figure \ref{fig:sigma_theta} one sees that this quantity follows a linear trend in the range of tilt we have investigated.
This trend seems to depend only slightly on \Ra. 
A fit of the straight line $1/\sigma_{\theta} = p + q\cdot \beta$ to the data revealed a slightly larger slope for the smaller \Ra\ ($q=67\pm 2$ for $\Ra=1.8\times 10^{10}$ and $q=61\pm2$ for $\Ra=7.2\times 10^{10}$).

\subsection{The oscillation of the LSC and its relation to the Reynolds number}

In addition to the random diffusion of the orientation and the amplitude of the LSC \cite[]{BA08a}, the LSC also performs a torsional oscillation when it is in the SRS [\cite{FA04}]. 
The orientations of a single roll oscillate with the same frequency near the top and the bottom of the roll, but with a phase shift of $\pi$ relative to each other. 
This motion is especially interesting since, for $\Gamma=1$, the oscillation period is the same as the turnover time of the single roll. Thus it can be used to calculate the Reynolds number \Rey\ of the flow \cite[]{QT02,BFA07,AGL09}.

Torsional oscillations were studied experimentally for $\Gamma=1$ \cite[]{BFA07,FBA08,XZZCX09}. They could be explained in terms of a stochastic model by \cite{BA09}.
For  $\Gamma=0.5$ there are stronger amplitude fluctuations and disruptions by flow-mode transitions than for $\Gamma = 1.00$ and thus these  oscillations are harder to detect experimentally.
Nevertheless, an analysis using only time segments for which the system was in the SRS and averaging over many of these revealed that  also for this $\Gamma$  a torsional mode exists \cite[]{WA11a}.  

As was done before \cite[]{FA04,ABN06,WA11a}, we uses the correlation functions 
\begin{equation}
\tilde C^{k_3,k_4}_{k_1,k_2}\equiv \langle \delta \theta_{k_1,k_2}(t) \delta \theta_{k_3,k_4}(t+\tau)\rangle
\label{eq:corr}
\end{equation}
of the differences between the orientations of the LSC at different heights in order to detect the torsional oscillations of the single roll. Here  $\langle \cdot\rangle$ indicates a time average, and  
\begin{equation}
\delta \theta_{k_i,k_j} \equiv \theta_{k_i} - \theta_{k_j}\mbox{,}
\end{equation}
where each of the $k_i$, $k_j$ ($i,j=1,\dots,4$) can be $``b"$, $``m"$, or $``t"$. The function given by Eq.~\ref{eq:corr} is then normalised according to 
\begin{equation}\label{eq:correl}
C_{k_1,k_2}^{k_3,k_4} = \frac{\tilde C^{k_3,k_4}_{k_1,k_2}(\tau)}{\sqrt{\tilde C_{k_1,k_2}^{k_1,k_2}(0)\cdot \tilde C_{k_3,k_4}^{k_3,k_4}(0)}} \mbox{.}
\end{equation}

\begin{figure}
\begin{center}
\includegraphics[width=0.95\textwidth]{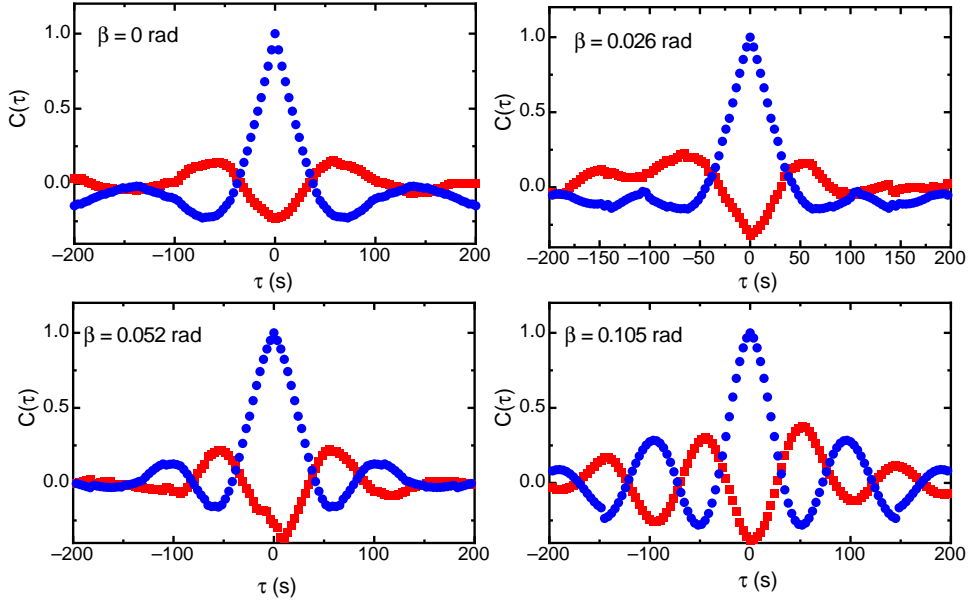}
\caption{The auto-correlation function $C^{t,b}_{t,b}$ (circles, blue online) and the cross-correlation function $C_{t,m}^{b,m}$ (squares, red online) as a function of the delay time $\tau$. Shown are correlation functions for different tilt angles $\beta$. All experiments were conducted at $\Ra=7.2\times 10^{10}$. }
\label{fig:SRM_Correl}
\end{center}
\end{figure}

We focus here on the auto-correlation function of the difference $\delta\theta_{t,b}$ and the cross-correlation function between the differences $\delta\theta_{t,m}$ and $\delta\theta_{b,m}$.  
As explained in detail by \cite{WA11a}, we calculated the correlation functions only  for the time-segments during which the system was in the SRS and then averaged over all such  segments.
Figure \ref{fig:SRM_Correl} shows the auto- and the cross-correlation functions ($C_{t,b}^{t,b}$ and $C_{t,m}^{b,m}$) for different tilt angles.
For the horizontal case, one can detect a clear anti-correlation for $\tau=0$ in $C_{t,m}^{b,m}$ and an oscillating behavior that manifests itself both in $C_{t,m}^{b,m}$ and $C_{t,b}^{t,b}$.
However, both correlation functions decay rather quickly, which is due to the erratic nature of the LSC associated with the randomly occurring  flow-mode transitions.

With increasing tilt angle, the twist oscillations of the correlation functions become more obvious  because the oscillation period becomes shorter and the  decay time becomes longer. In addition, the correlation functions become smoother because the lengths of the SRS-segments in the time series increase with increasing tilt angle (see Sec. \ref{sec:SRS-DRS}).

\begin{figure}
\begin{center}
\includegraphics[width=0.95\textwidth]{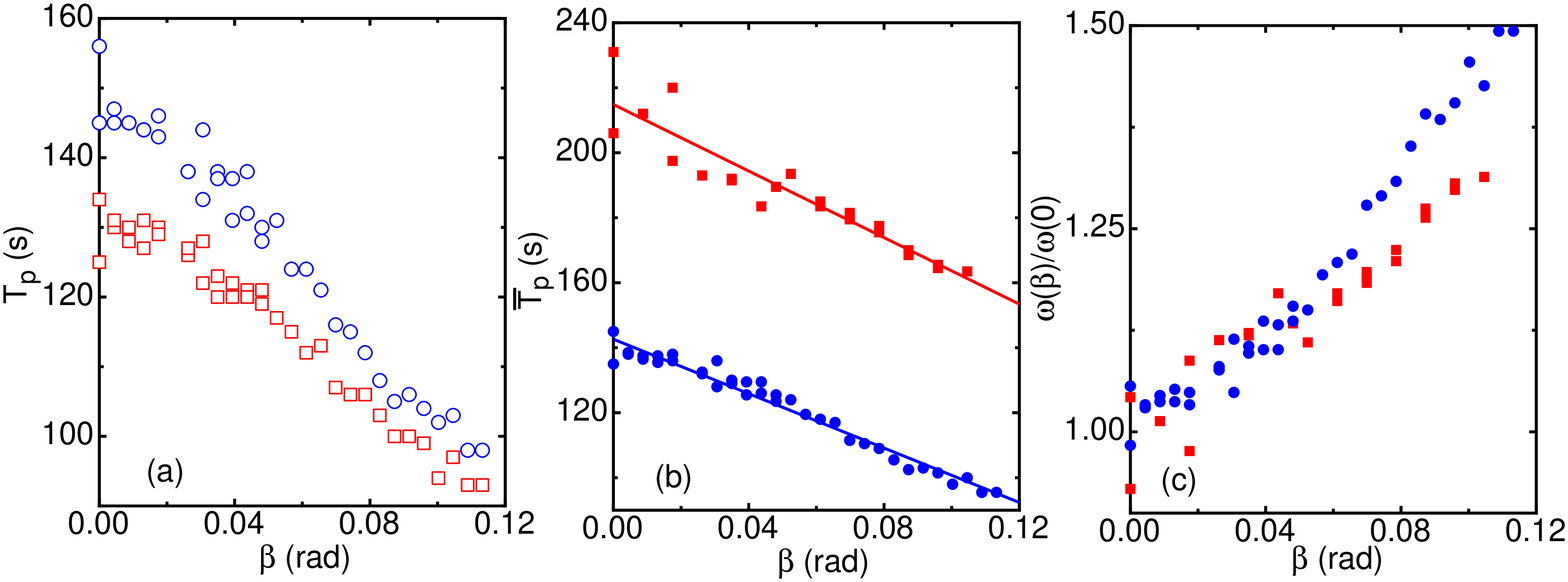}
\caption{(a): Torsional periods as measured from $C_{t,m}^{b,m}$ (open squares, red online) and  $C_{t,b}^{t,b}$ (open circles, blue online) for $\Ra=7.2\times 10^{10}$. 
(b): Averages $\overline{T}_p$  of the torsional periods obtained by the two methods  for $\Ra = 1.8\times 10^{10}$ (solid squares, blue online) and  $\Ra=7.2\times 10^{10}$ (solid circles, blue online).
(c): Normalised torsional frequency $\omega(\beta)/\omega(0)$ for $\Ra=1.8 \times 10^{10}$ (solid squares, red online) and for $\Ra=7.2\times 10^{10}$ (solid circles, blue online). }
\label{fig:twist}
\end{center}
\end{figure}

It is desirable  to quantify the change of the torsional oscillation-frequency with increasing $\beta$. A fit of an appropriate oscillating but decaying function to the data turned out to be difficult because of the rapidly decaying envelope and the experimental irregularities of the functions that are caused by the relatively short time periods over which the SRS was available, particularly at the smaller $\beta$.  
Thus, we used two ways to determine the oscillation period $T_p$. We measure the distance between the two maxima closest to $\tau=0$ in the cross-correlation function $C_{t,m}^{b,m}$, and we determine the distance between the first two minima in the auto-correlation function $C_{t,b}^{t,b}$. Both values are not exactly the same because of the decaying envelope which shifts the maxima (minima) slightly towards smaller values of $\tau$. 
The results  obtained this way are shown in figure \ref{fig:twist}a.
One sees that the periods calculated from $C_{t,b}^{t,b}$ (bullets, blue online) are slightly larger than those determined from $C_{t,m}^{b,m}$ (squares, red online). However, both follow the same trend, namely a decrease of the period with increasing tilt.

In Fig. \ref{fig:twist}b we display the averages $\overline{T}_p$   of the torsional periods obtained by the two methods  for $\Ra = 1.8\times 10^{10}$ (solid squares) and  $\Ra=7.2\times 10^{10}$ (solid circles) as a function of $\beta$. They can be represented quite well by the straight lines $\overline{T}_p = r + s\cdot \beta$, as shown by the lines in the figure. Least-squares fits yielded $r = 143~ (215)$ s and $s = -419~§ (-511)$ s/rad for $\Ra = 7.2\times 10^{10} ~(\Ra=1.8\times 10^{10})$. 

Figure \ref{fig:twist}c shows the normalised oscillation frequencies $\omega(\beta)/\omega(0) = \overline{T}(0)/\overline{T}(\beta)$. 
We normalised the results in order to compare the increase of the frequency for $\Ra=1.8\times 10^{10}$ with the one for $\Ra=7.2\times 10^{10}$.
The $\beta$-dependence of the normalised frequency is similar for both \Ra; however, there is a slightly smaller increase for the smaller \Ra. 
If we assume that the oscillation frequency is proportional to the average velocity of the flow ({\em i.e.} the Reynolds number \cite[]{QT02}), then the results here should be comparable with the increase of the LSC amplitude $\delta_k$ as discussed in Sec. \ref{sec:delta}. Although there is a qualitative similarity, one sees that $\delta(\beta)/\delta(0)$ increases more rapidly with $\beta$ than does $\omega(\beta)/\omega(0)$, by about a factor of three or so.
Another difference is that for $\omega(\beta)/\omega(0)$ the data for the smaller \Ra\ show a slightly smaller increase in comparison with the large-\Ra\ data; for $\delta(\beta)/\delta(0)$ it is the other way around. The reasons for these differences are unclear. On the one hand the mapping between the azimuthal temperature variation that yields $\delta$ and the velocity field may not be as simple as assumed. On the other hand, the LSC oscillation frequency for $\Gamma = 0.50$ may not simply be proportional to the Reynolds number of the SRS.

 \section{Conclusion}

In this paper we reported on an investigation of the effect of a small tilt on the heat transport and the flow structure in turbulent Rayleigh-B\'enard convection (RBC) in a cylinder with aspect ratio $\Gamma=0.50$.
In particular we showed results for tilt angles up to $\beta=0.122$ rad and for the two Rayleigh numbers $\Ra=1.8\times 10^{10}$ and $\Ra=7.2\times 10^{10}$.

We found that a small tilt enhanced the heat transport very slightly. This differs from the case of RBC in cylindrical samples  with aspect ratio $\Gamma=1.0$ where a very small decrease had been observed. 
By looking at the ratio of the times that the system spends in the single-rolls state (SRS) and the double-roll state (DRS), we found that the SRS is stabilized by the tilt and the time that the system spends in the DRS is reduced. For $\beta \geq 0.06$ the SRS is found essentially all the time.
Since the SRS transports heat more effectively than the DRS, its stabilization explains the slight increase of the Nusselt number \Nu\ with increasing tilt.

To investigate the behavior of the large-scale circulation (LSC) in the presence of a tilt, we measured the average amplitude of its first Fourier mode ($\delta_k$) and found an increase with increasing tilt.
The relative increase was slightly larger for smaller \Ra.
A stabilisation of the flow could be seen as well in the event rate, which is the frequency at which the amplitude drops below a critical value taken to be 15\% of its average value.
Decreasing numbers of events were found with increasing tilt. For tilt angles larger 0.06\,rad no events could be detected during the experimental run time of a day for each data point.

Also the dynamics of the LSC orientation $\theta_k$ was influenced by the tilt. The orientation was concentrated in one direction and random diffusion over all angles was suppressed. This could be seen in the probability density function of $\theta_k$ which became narrower. Over the investigated range of tilt angles, the inverse of the standard deviation ($1/\sigma_\theta$) of $p(\theta_k)$ was proportional to the tilt angle. 
The torsional oscillation at which the top part of the single roll oscillates relative to the bottom part became more pronounced and the oscillation frequency increased with increasing tilt angle.
For $\Gamma = 1.00$ the frequency had been shown to be a measure for the Reynolds number. Thus its increase indicates an increasing flow velocity.

The stabilisation of the LSC can also be seen in a Fourier analysis of the first 4 modes that can be measured from the azimuthal temperature profile along the sidewall. With increasing tilt the relative energy in the first mode increases at the expense of the others. 
While in the horizontal case the relative energy in the first mode is smaller for smaller \Ra, for increasing tilt the energy of the first mode becomes nearly independent of \Ra.

\begin{acknowledgments}
This work was supported by the U.S. National Science Foundation through Grant DMR-1158514.
\end{acknowledgments}



\end{document}